\documentclass[12pt, twoside]{article}
\usepackage[a4paper]{geometry}
\usepackage{tgpagella}
\usepackage{graphicx}
\usepackage{epigraph}

\usepackage{comment}
\usepackage{physics}
\usepackage{microtype}
\usepackage{amsmath}
\usepackage[bottom]{footmisc}
\usepackage{footmisc} 

\usepackage{titlesec} 
\titleformat{\section}[runin]
        {\normalfont\bfseries}
        {\thesection.}
        {0.5em}
        {}
        [.]
        
        \titleformat{\subsection}[runin]
        {\normalfont\itshape}
        {\thesubsection.}
        {0.5em}
        {}
        [.]
        
\titlespacing*{\section}{0pt}{5mm}{2.5mm}

\titlespacing*{\subsection}{\parindent}{2.5mm}{3mm} 

\usepackage{fancyhdr} 
\fancypagestyle{plain} 
{
\fancyhead[LCR]{} 
\fancyfoot[C]{\thepage}
   
}

\usepackage{graphicx, listings, setspace, hyperref, verbatim} 
\usepackage{amsmath, amssymb, mathtools, esint}

\usepackage[T1]{fontenc}
\usepackage[utf8]{inputenc}
\usepackage{mathptmx}

\usepackage{filecontents}

\usepackage[backend=bibtex, authordate, url=false, doi=false, eprint=false]{biblatex-chicago} 
\AtEveryBibitem{%
  \clearfield{note}%
  \clearfield{series}%
  \clearfield{pagetotal}%
  \clearfield{number}
  \clearfield{chapter}%
}

\DeclareFieldFormat{postnote}{\mkcomprange{#1}} 

\DeclareFieldFormat{pages}{\mkcomprange{#1}} 

\DefineBibliographyStrings{english}
{ byeditor = {ed.}, } 

\hypersetup{
	pdftitle={Your Paper Title},
	pdfauthor={},
	pdfsubject={},
	pdfkeywords={},
	pdfcreator={XeLaTeX},
	pdfproducer={XeLaTeX},
	pdftoolbar=false,	
	pdfmenubar=true,	
	pdffitwindow=false,	
	pdfstartview={FitH},	
	unicode=true,		
	pdfnewwindow=true,	
	colorlinks=true,	
	linkcolor=blue,	
	citecolor=black,	
	filecolor=black,	
	urlcolor=blue		
}

\newif\iffootpunct 

 \makeatletter
  
 \def\blfootnote{\gdef\@thefnmark{} \@footnotetext} 
 
 \renewcommand{\@makefntext}[1]{%
  \setlength{\parindent}{0pt}%
  \begin{list}{}{\setlength{\labelwidth}{0mm}
    \setlength{\leftmargin}{\labelwidth}%
    \setlength{\labelsep}{0pt}
     \setlength{\itemsep}{0pt}%
      \setlength{\parsep}{0pt}%
      \setlength{\topsep}{0pt}
    \singlespacing \footnotesize \vspace{-3.5 ex}}
  \item[\@thefnmark \iffootpunct. \fi \hfil ]#1
  \end{list}%
}

\makeatother

\usepackage[main=american]{babel}
\usepackage[babel=true]{csquotes} 

\begin{document}

\title{\vspace{-0.5em} \LARGE T Falls Apart: \\ On the Status of Classical Temperature in Relativity\vspace{1em}}			
\author{\Large Eugene Y. S. Chua\textsuperscript{*\textdagger‡}}
\date{\vspace{-8.1ex}}
\maketitle

\vspace{5mm}

\noindent\rule{\textwidth}{1.5pt}

\noindent {\small{Taking the formal analogies between black holes and classical thermodynamics seriously seems to first require that classical thermodynamics applies to relativistic regimes. Yet, by scrutinizing how classical temperature is extended into special relativity, I argue that it falls apart. I examine four consilient procedures for establishing classical temperature: the Carnot process, the thermometer, kinetic theory, and black-body radiation. I show how their relativistic counterparts demonstrate no such consilience in defining relativistic temperature. Hence, classical temperature doesn’t appear to survive a relativistic extension. I suggest two interpretations for this situation: eliminativism akin to simultaneity, or pluralism akin to rotation.}}

\noindent\rule{\textwidth}{1.5pt}

\vspace{0mm}

\footpunctfalse

\blfootnote{*To contact the author, please write to: Department of Philosophy, 9625 Scholars Drive North, 4th Floor, La Jolla, CA 92093-0119, USA; e-mail: eychua@ucsd.edu.}
\blfootnote{\textdagger This paper was made possible through the support of a grant from the John Templeton Foundation. The opinions expressed in this paper are those of the author and do not necessarily reflect the views of the John Templeton Foundation. Many thanks to Craig Callender, Samuel Fletcher, Bosco Garcia, Nick Huggett, Sai Ying Ng, John Norton, Mike Schneider, Ann C. Thresher, attendees of the 2022 AGPhil DPG conference,  for helpful comments and suggestions. Thanks also to Thomas Barrett for kindly sharing his undergraduate dissertation on this topic with me.}
\blfootnote{‡This is a preprint version of a paper which has been accepted at \textit{Philosophy of Science}.}


\footpuncttrue

\setlength{\parindent}{1em} 

\epigraph{\footnotesize{Dear Laue: I hear the voice of my conscience when I remind you of the dispute concerning the rendering of the fundamental thermodynamic concepts in the special-relativistic form. There is actually no compelling method in the sense that one view would simply be 'correct' and another 'false'. One can only try to undertake the transition as naturally as possible.}}{\footnotesize{Albert Einstein, \\1953 letter to Max von Laue}}

\section{Introduction}

Do the laws and concepts of classical thermodynamics (\textbf{CT}) hold a universal character? Einstein, for instance, wrote that ``[\textbf{CT}] is the only physical theory of universal content concerning which I am convinced that, within the framework of the applicability of its basic concepts, it will never be overthrown.” (1946/1979, 33) Given such proclamations, and how research in black hole thermodynamics -- birthed by formal analogies with \textbf{CT} -- continues to this day, one naturally assumes that \textbf{CT} \textit{can} be extended into the relativistic regime and beyond -- there is no limit to the ``framework of applicability" of its basic concepts.\footnote{For more on black hole thermodynamics and its formal analogies with thermodynamics, see e.g. Bekenstein's (1973 and 1975). See Dougherty $\&$ Callender (2016) for criticism, and Wallace (2018) for a rejoinder.}   

It is therefore interesting that a parallel debate drags on without resolution. Although Planck and Einstein pioneered the special relativistic extension of thermodynamical concepts by developing a set of Lorentz transformations, they by no means settled the issue. Importantly, \textit{temperature} resists a canonical relativistic treatment: there are different equivocal ways of relativizing temperature. While physicists appear to treat this as an empirical problem,\footnote{For instance, Farias et al (2017) remarks that ``the long-standing controversy [...] is mainly based on the initial assumptions, which need to be tested [...] to discern which set of Lorentz transformations is correct for quantities such as temperature". } or something to be settled by convention,\footnote{Landsberg $\&$ Johns (1967) suggests that the choice of Lorentz transformation for temperature could be ``settled by convention".} the issue seems \textit{conceptually} problematic to me.

I argue that this situation suggests a breakdown of the classical non-relativistic concept of temperature -- T$_{classical}$ -- in special-relativistic regimes, i.e. when we consider the temperature of a relatively moving body at high speeds. The procedures which jointly provided physical meaning to T$_{classical}$ do not do so in relativistic settings. T$_{classical}$ breaks down in this regime; there \textit{is} a limit to the framework of applicability of the \textit{classical} thermodynamic concepts. 

Notably, my argument will rest \textit{not} on the fact that there is no way of defining temperature in relativistic regimes, but that there are \textit{many}, \textit{equally valid} procedures for defining the relativistic temperature which \textit{disagree} with each other. I focus on four procedures:

\begin{itemize}
    \item one can attempt to construct a relativistic Carnot cycle,
    \item use a co-moving thermometer,
    \item consider a relativistic extension of kinetic theory and particle mechanics,
    \item or scrutinize the black-body radiation of a moving body.
    
\end{itemize}

\noindent I chose these four because their classical counterparts were significant in determining the physical meaning of T$_{classical}$: its theoretical relationship with heat (via the Carnot cycle), its phenomenology (with a thermometer), its ontology (via particles), and its connection with radiation (via black-body radiation). It is through this lens that I propose we understand Einstein's above notion of `natural'-ness: there is strong \textit{consilience} between these procedures, in the operational sense that the temperature established via any of these procedures agrees with the temperature in other procedures.\footnote{Given a proper understanding of the approximation, idealization, and de-idealization procedures.} Contrariwise, their relativistic counterparts demonstrate no such consilience: different procedures predict starkly different behaviors for the temperature of a moving body. Furthermore, for each procedure, we find conceptual difficulties too. `Natural' procedures in \textbf{CT} -- which generated a consilient and robust concept of temperature -- do not appear to be `natural' at all in relativistic settings. 

I end by proposing two possible interpretations of this situation: an eliminativist one on which we interpret temperature akin to simultaneity, or a pluralist one on which we interpret temperature akin to relativistic rotation. 

\section{The Quest to Relativize Thermodynamics: The Odd Case of Temperature} 

I focus on attempts to relativize \textbf{CT},\footnote{I refer to the usual classical / phenomenological set of laws governing a system's approach to equilibrium, the meaning of equilibrium, conservation of energy in terms of heat and work, entropy non-decrease, and entropy at the absolute zero of temperature. See e.g. Planck (1945) for a \textit{locus classicus} on the topic.} i.e. some extension of its laws and concepts into \textit{special} relativity.\footnote{That is, we assume that events occur on a background Minkowski (flat) spacetime with signature $\{-, +, +, +\}$, where the allowed coordinate frames are inertial frames, that is, frames or observers moving at constant velocity (or zero velocity).} 

The pioneers of relativistic thermodynamics, e.g. Einstein (1907) and Planck (1908), sought a set of Lorentz transformations for the laws and quantities of \textbf{CT},\footnote{The details are excellently summarized in Liu (1992 and 1994).} just as we have for e.g. position and time. For instance, an observer $O'$ (or the associated inertial frame) with positions and times ($x'$, $y'$, $z'$, $t'$) moving along the $x$-axis away from another observer $O$ (and their inertial frame) at constant velocity $v$ can be understood by $O$ to be at positions and time ($x$, $y$, $z$, $t$) via:

\begin{equation}
\begin{array}{l}
    t' = \gamma(t - \frac{vx}{c^2}) \\
    x' = \gamma(x - vt) \\
    y' = y \\
    z' = z \\
\end{array}
\end{equation}

\noindent where $\gamma = \frac{1}{\sqrt{1 - \frac{v^2}{c^2}}}$ is the Lorentz factor, and $c$ is the speed of light. 

Relativistic thermodynamics hopes to find similar transformations for thermodynamic quantities like temperature, pressure, volume, etc. The underlying assumption is that thermodynamics can be shown to have physical meaning in relativistic regimes only when we have a set of Lorentz transformations under which thermodynamic quantities can be shown to transform, just as we do for position and time.\footnote{That the only physically meaningful quantities are ones which are invariant or covariant under Lorentz transformations, and that the laws must hold true in similar fashion in all inertial frames, is a common idea in relativity. See Lange (2002, 202) or Maudlin (2011, 32) for an exposition of this idea.}

Planck and Einstein successfully derived the transformations for most thermodynamic quantities like pressure $p$, volume element $dV$, and entropy $S$:\footnote{The argument for entropy's Lorentz invariance is generally accepted; I will do the same here. However, see Earman (1986, 177--178) and Haddad (2017, 41 -- 42) for criticisms of Planck's argument.} 

\begin{equation}
\begin{array}{l}
    dV' = \frac{dV}{\gamma} \\
    p' = p \\
    S' = S \\
\end{array}
\end{equation}

\noindent Fixing $S$ appears to indirectly fix the concepts of heat and temperature, via the well-known relation $Q = TdS$. However, surprisingly, the Lorentz transformation for \textit{temperature} turns out to be highly equivocal. 

\section{The Classical Temperature}

In \textbf{CT}, at least four well-known procedures exist for establishing the concept of temperature. Notably, there is significant \textit{consilience} between them, which suggests there really is a physically significant quantity: T$_{classical}$.

\subsection{The Carnot Cycle}

The Carnot cycle is a foundational theoretical concept in \textbf{CT} by which we can define absolute temperature in terms of heat.\footnote{For an excellent historical account of Lord Kelvin's definition of the classical temperature via the Carnot cycle, see Chang (2004, Ch. 4).} The typical idealized example is an ideal gas acting on a piston in a cylinder (the `engine') while undergoing reversible processes (see Figure 1):

\begin{enumerate}
    \item The gas receives heat $Q_2$ from a heat bath at temperature $T_2$ and isothermally expands, doing work on the surroundings.
    \item The cylinder is thermally insulated, and the gas adiabatically expands and continues to do work on the environment, decreasing in temperature to $T_1$.
    \item The gas is isothermally compressed at $T_1$ at the second heat bath, losing  $Q_1$ to the heat bath.
    \item The cylinder is thermally insulated, and the gas is adiabatically compressed as the environment continues to do work on the gas. 
    \item The cylinder is then brought back to the initial heat bath with $T_2$.
\end{enumerate}

\begin{figure}[ht]
\includegraphics[scale = 0.40]{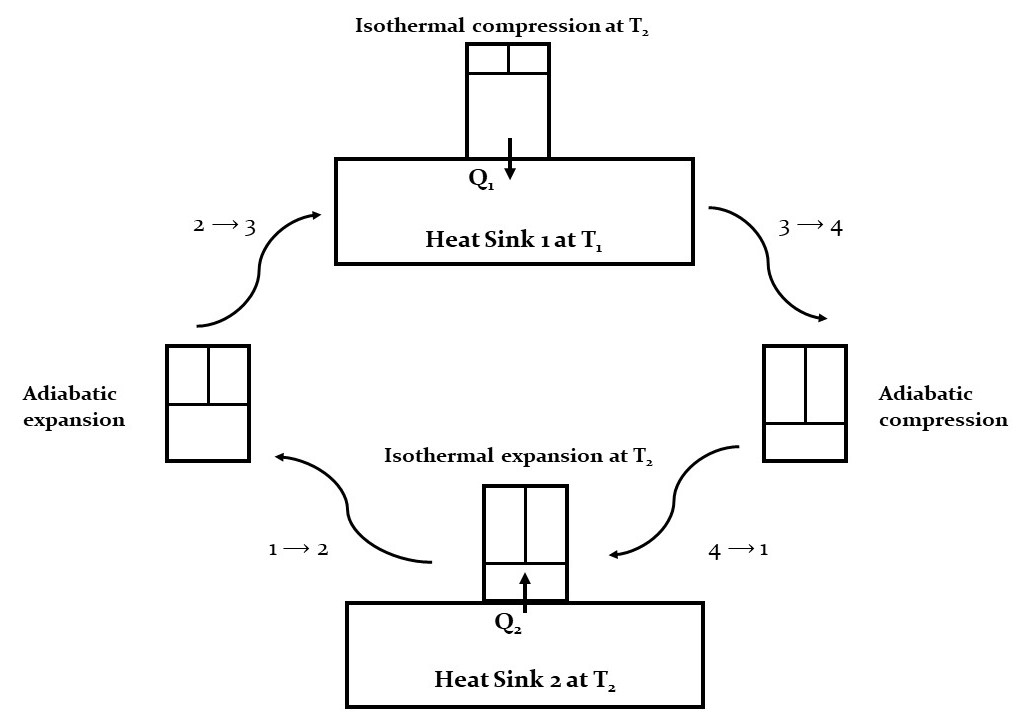}
\centering
\caption{An example of a Carnot cycle, with $T_2 > T_1$.}
\end{figure}

\noindent In such a cycle, a simple relationship between the heat exchange to and from a heat reservoir, and their temperature, can be derived. In a foundational statement of the relationship between heat and temperature in classical thermodynamics, Joule and Thomson (1854/1882) wrote:

\begin{quote}
If any substance whatever, subjected to a perfectly reversible cycle of operations, takes in heat only in a locality kept at a uniform temperature, and emits heat only in another locality kept at a uniform temperature, the temperatures of these localities are proportional to the quantities of heat taken in or emitted at them in a complete cycle of operations. [say, a Carnot cycle laid out above] (1854/1882, 394)
\end{quote}

\noindent Formally, it is a remarkably simple statement: 

\begin{equation}
    \frac{T_1}{T_2} = \frac{Q_1}{Q_2}
\end{equation}

\noindent If one can calculate the amount of heat exchanged between the two reservoirs, we can then theoretically derive the ratio between the two reservoirs' temperatures. With this in hand, Thomson, as McCaskey (2020, 32) puts it, proposed that ``we should construct a temperature scale from the mathematics of a non-existent idealized Carnot engine and the behaviour of a non-existent idealized gas. Any differences between the calculated temperature and the readings of a thermometer will be attributed to shortcomings in the thermometer."

This is the problem: how do we actually connect this to actual measurements and the world? Actual Carnot cycles are hard to construct realistically, and they depend crucially on the fact that gases in these cycles obey the ideal gas law:

\begin{equation}
    PV = nRT
\end{equation}

\noindent where $T$ is the temperature defined by (3).\footnote{$P$ is pressure, $V$ is volume, $n$ the amount of substance, and $R$ the ideal gas constant.} While no gases are strictly ideal gases, Thomson,\footnote{For a discussion of Thomson's strategies and the extent to which they succeeded, see Chang $\&$ Yi (2005).} and later Callendar (1887) and Le Chatelier $\&$ Boudouard (1901), pursued a variety of operationalizations and experiments which succeeded in measuring the extent to which actual gases deviated from the predicted behavior of ideal gases, i.e. deviated from (4): only about $0.628^{\circ}$ at 1000$^{\circ}$C for constant-volume air thermometers, and $1.198^{\circ}$ for constant-pressure air thermometers. In other words, actual air thermometers approximated theoretical ideal gas thermometers remarkably well. In doing so, we can establish how measurements of temperature derived from observations of actual gases approximates that of the theoretical temperature predicted for ideal gases in Carnot cycles. This, in turn, establishes the legitimacy of the theoretical notion of classical temperature defined here.\footnote{For a much more detailed discussion of just how complicated these operationalizations were, and why they are nevertheless successful, see Chang (2004, sub-section ``Analysis: Operationalization—Making Contact between Thinking and Doing").}

\subsection{The Thermometer}

This brings us to thermometers and how they establish the concept of temperature. One crucial development was Fahrenheit's invention of a reliable thermometer, which allowed one to make independent measurements of this physical quantity called the temperature, consistently compare said measurements, and grasp temperature as a robust and (fairly) precisely measured numerical concept rather than one associated with vague bodily sensations. It is not an understatement to say that the classical notion of temperature would not be developed without such an invention.\footnote{For an involved discussion of this development, see Chang (2004, chs. 1 and 2) For a general overview, see McCaskey (2020).}

While thermometers made with the \textit{same} material \textit{were} reliable with respect to each other, thermometers made with \textit{different} materials differed in their rates of expansion and contraction. Importantly, the Carnot cycle discussed above provides a \textit{theoretical} foundation for temperature by providing a definition of temperature independent of material. As Thomson himself explained: 

\begin{quote}
    As reference is essentially made to a specific body as the standard thermometric substance, we cannot consider that we have arrived at an absolute scale...
\end{quote}

\noindent In contrast:

\begin{quote}
    The relation between motive power and heat, as established by Carnot, is such that quantities of heat, and intervals of temperature, are involved as the sole elements in the expression for the amount of mechanical effect to be obtained through the agency of heat; [...], we are thus furnished with a measure for intervals according to which absolute differences of temperature may be estimated. (Thomson 1848/1882, 102)
\end{quote}

\noindent In other words, the Carnot cycle is intended to provide theoretical foundations to the observed measurements of temperature provided by actual thermometers. 

However, as we have already seen, actual thermometers themselves were in turn crucial for supporting the theoretical notion of temperature. Given the difficulties in building an actual Carnot engine and the non-ideal nature of actual gases, actual air thermometers provided a means of de-idealization. They showed how actual temperature measurements can be understood as approximating the abstract theoretical notion of temperature defined by an idealized Carnot cycle and ideal gases. 

Put another way, in yet another sign of consilience, we can understand the concept of classical temperature to be co-established by both the observed quantity of temperature provided by a thermometer \textit{and} the theoretical quantity provided by the Carnot cycle, with each procedure supporting the other. 

\subsection{Kinetic Theory of Heat}

The kinetic theory of heat provides another way to understand temperature via the Maxwell-Boltzmann distribution. For a system of ideal gas\footnote{Here an ideal gas is interpreted as a set of $n$ identical weakly interacting particles.} in equilibrium with temperature $T$, the Maxwell-Boltzmann distribution connects the notion of temperature explicitly with the notion of bulk particle velocities via:

\begin{equation}
    f(v) = \sqrt{\left(\frac{m}{2\pi kT}\right)^3}4\pi v^2 e^{-[\frac{1}{2}mv^2 + V(x)]/{kT}}
\end{equation}

\noindent where $m$ is the particle's mass, $T$ is the temperature, $k$ is Boltzmann's constant, V(x) is the system's position-dependent potential energy, and $v$ is an individual molecule's velocity.\footnote{For a historical account of this distribution, see Brush (1983, $\S1.11$).} Importantly, $f(v)$ tells us, for a system of ideal gas in equilibrium, how many particles we expect to find with some range of velocities $v$ to $v + dv$, given some temperature.

\begin{figure}[ht]
\includegraphics[scale = 0.24]{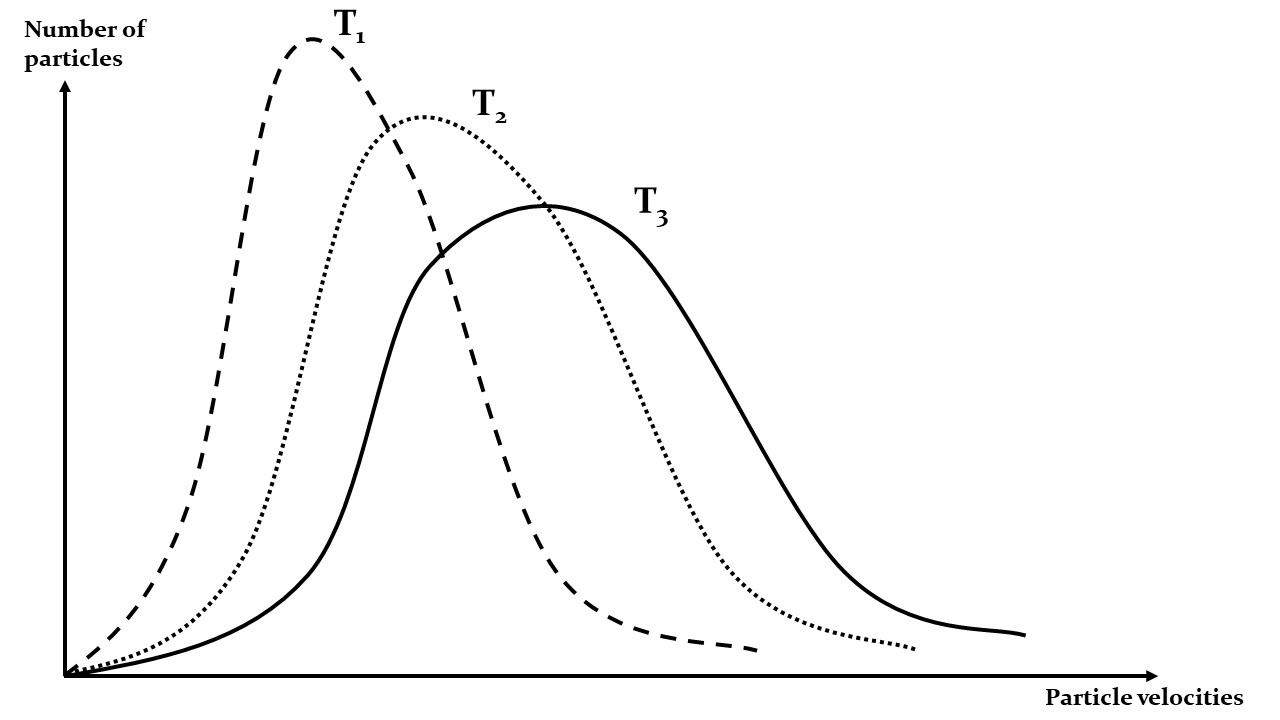}
\centering
\caption{A schematic Maxwell-Boltzmann distribution at three different temperatures $T_1 < T_2 < T_3$.}
\end{figure}

Figure 2 shows the schematic connection between a gas's temperature and velocities of the particles composing said gas. This distribution plays a significant conceptual role by allowing us to derive the well-known relationship between temperature and mean kinetic energy: 

\begin{equation}
    \langle \frac{1}{2}mv^2 \rangle = \frac{3}{2}kT
\end{equation}

\noindent This provides foundational support for thermodynamics -- and the concept of temperature -- in terms of particle mechanics, by allowing us to understand the concept of a system's temperature in terms of the mean kinetic energy of particles composing said system.\footnote{This has been much discussed in the philosophical literature. See e.g. De Regt (2005) and references therein.}

Importantly, though, this distribution holds only for ideal gases. As such, the de-idealization of ideal gases in terms of actual gases, already mentioned in previous sections, also plays a role here and further highlights the consilience of these various procedures in establishing the concept of temperature. They show us that measured temperatures of actual gases are approximately the temperatures of ideal gases with similar pressures and volumes; in turn, the idealized relationship between mean kinetic energy and temperature here, which holds for ideal gases, can also be understood to approximate that of real systems of particles.\footnote{In particular, highly dilute gases at non-extremal temperatures.} This provides us with a reason to accept that the temperature of actual systems really can be understood in terms of mean kinetic energy as well, providing further physical meaning to the concept of temperature. There is, again, a consilience between the concept of temperature employed here and the concept of temperature developed through the other procedures. 

\subsection{Black-Body Radiation}

Finally, the study of black-body radiation connects temperature to electromagnetic radiation. A black-body is defined as one which absorbs (and emits) all thermal radiation incident upon it without reflecting or transmitting the radiation, for \textit{all} wavelengths and angles of incident upon it. Notably, since a black-body does not distinguish directionality, it emits \textit{isotropic} radiation.

There are simple laws relating the properties of radiation to the black-body's temperature.\footnote{For a historical account, see Brush (1983, $\S3.1$) or Stewart $\&$ Johnson (2016).} Firstly, the Stefan-Boltzmann law states:

\begin{equation}
    j^* = \sigma T^4
\end{equation}

\noindent where $j^*$ is the total energy emitted per unit surface area per unit time by the black-body, $T$ is its temperature, and $\sigma$ is the Stefan-Boltzmann constant.\footnote{For less idealized bodies which do not absorb all radiation, the law is given by:

\begin{equation}
    j^* = \epsilon \sigma T \nonumber 
\end{equation}

where $0 < \epsilon < 1$ is the emissivity of the substance.}

Secondly, Wien's displacement law:

\begin{equation}
    \rho(f, T) = f^3 g(\frac{f}{T})
\end{equation}

\noindent states that the energy density $\rho$ of radiation from systems with temperature $T$, at frequency $f$, is proportional to $f^3 g(\frac{f}{T})$ for some function $g$.\footnote{See Brush (1983, Ch. 3) for a historical narrative.}  Integrating over all possible $f$ amounts to computing the total energy density of radiation from all frequencies, and entails the Stefan-Boltzmann law regardless of choice of $g$. Furthermore, if $\rho(f, T)$ achieves its maximum for some value of $f$, $f_{max}$, then:

\begin{equation}
    f_{max} \propto T
\end{equation}

\noindent or in terms of peak wavelength $\lambda_{peak}$:  

\begin{equation}
    \lambda_{peak} \propto \frac{1}{T}
\end{equation}

\noindent This captures the familiar observation that things which are heated first turn red and then into other colors associated with higher frequencies -- and hence shorter wavelengths -- as their temperature increases.

These laws form the foundations of the relationship between radiation and temperature for a radiating black-body in equilibrium in \textbf{CT}. In particular, much of the work in the late 19\textsuperscript{th} century revolved around the search for the appropriate function $g$, and modifications or generalizations to Wien's law, pursued by others like Rayleigh, Planck, and Einstein. 

Interestingly, as another mark of consilience, Einstein's famous 1905 discussion of the quantization of thermal radiation drew upon analogies between thermal radiation and the ideal gas law as well as the Maxwell-Boltzmann distribution for ideal gases: the energy density formula one extracts from the former looks remarkably like the latter.\footnote{See Norton (2005) for a discussion of the analogies and disanalogies between the two. See also Uffink (2006). See Fowler (n.d., ``Einstein Sees a Gas of Photons") for a preliminary exposition of the analogy.}  This led Einstein to conclude that we can understand thermal radiation as quantized, analogous to how an ideal gas can be understood as composed of a number of particles. Crucially for our purposes, we once again see how the same classical temperature concept is supported and applied across different procedures, establishing temperature as a physically significant and meaningful concept across these contexts. 

\section{From Classical to Relativistic Temperature}

In \textbf{CT}, the above procedures show remarkable consilience in that the very same concept of temperature can be determined or understood in terms of any of these procedures without much practical issue. For instance, the temperature observed by a thermometer for a radiating body is approximately the temperature deduced via the observed frequencies of their radiation. A box of gas can equilibrate with a radiating system and both will come to the same temperature.\footnote{Einstein's famous 1905 discussion of the quantization of thermal radiation drew upon analogies between thermal radiation and the Maxwell-Boltzmann distribution: the energy density formula one extracts from the former looks remarkably like the latter. This led Einstein to conclude that we can understand thermal radiation as quantized, analogous to how an ideal gas can be understood as composed of a number of particles. See Norton (2005) and Uffink (2006) for discussion. See Fowler (n.d., ``Einstein Sees a Gas of Photons") for a preliminary exposition of the analogy.} Finally, the theoretical definition of temperature found by considering a Carnot cycle can also be connected back to the empirical temperature measurements of actual thermometers.

This consilience then motivates why we might find the concept of temperature -- and its application in these contexts -- `natural', to borrow Einstein's words. After all, these procedures clearly refer to some quantity which can be measured, manipulated, understood, compared, and calculated across various contexts.

However, there is no such consilience when attempting to relativize temperature. Each procedure establishes a \textit{different} notion of relativistic temperature, and not without conceptual difficulties.

\subsection{The Relativistic Carnot Cycle: Moving Temperature is Lower/Higher}

I begin with the relativistic Carnot cycle.\footnote{See Liu (1992), Liu (1994) for a detailed historical overview of the topic. See Farias et al (2017) for a physics-oriented overview, and Haddad (2017, 39 -- 42) for a concise overview of the disagreements on this procedure.} Von Monsengeil, who devised the process, explicitly appealed to the foundational role of the classical Carnot cycle in defining T$_{classical}$, and proposes an extension to that procedure. (von Mosengeil 1907, 160 -- 161) Essentially we demand that the same relations between heat and temperature hold when one heat bath is now \textit{moving} with respect to the other with some velocity $v$ (see Figure 3). We are supposed to adiabatically accelerate the engine (i.e. the piston and cylinder of gas) from one inertial frame to another, and adiabatically decelerate it back to the original frame in completing this cycle. 

Just as a classical Carnot cycle defined a relationship between the temperature and heat exchange of two heat baths, a relativistic Carnot cycle is stipulated to do the same. For a heat bath at rest with temperature $T_0$ and another moving with respect to it with a `moving temperature' $T'$, with the engine \textit{co-moving} with the respective heat baths during the isothermal processes:

\begin{equation}
    \frac{T'}{T_0} = \frac{Q'}{Q_0}
\end{equation}

\noindent and hence:

\begin{equation}
    T' = \frac{Q'}{Q_0}T_0
\end{equation}

\noindent What remains is `simply' to define the appropriate heat exchange relations. That turns out precisely to be the problem: there are two ways to understand the heat exchange between the engine and the moving heat bath, and there does not seem to be a fact of the matter which is appropriate.\footnote{See Liu (1992) for a much more detailed discussion of this disagreement.}

\begin{figure}[ht]
\includegraphics[scale = 0.40]{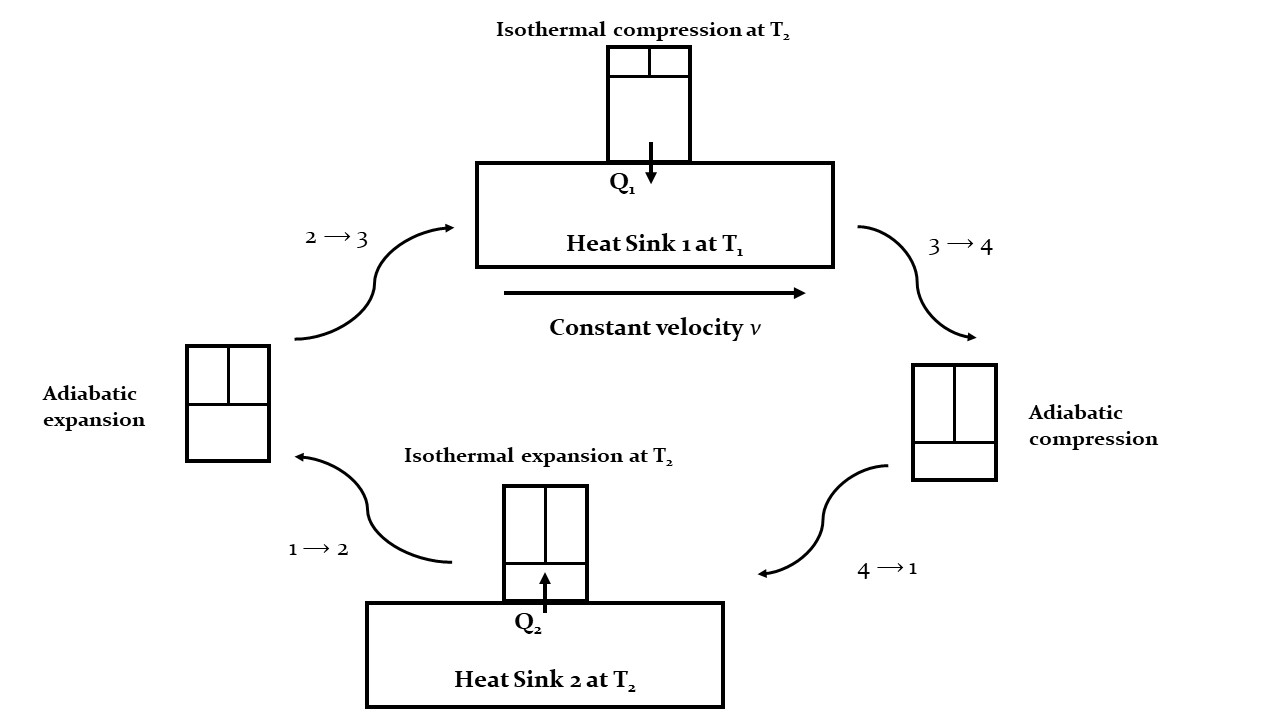}
\centering
\caption{A relativistic Carnot cycle.}
\end{figure}

Firstly, one may, like Planck and early Einstein, understand the heat transfer from the perspective of the rest frame of the stationary bath. Then the engine, in exchanging heat with the moving bath, also exchanges energy. By relativistic mass-energy equivalence, this causes the bath to lose or gain momentum by changing its mass. However, without further work, the bath then cannot stay in inertial motion - it will decelerate or accelerate. Hence, to keep it moving inertially, we need to perform extra work on it, which Einstein proposed to be:

\begin{equation}
    dW = pdV - \textbf{u} \cdot d\textbf{G}
\end{equation}

\noindent $pdV$ is simply the usual compressional work done by the piston due to the gain or loss of heat from the bath. However, there is a crucial inclusion of the $\textbf{\textbf{u}} \cdot d\textbf{G}$ term, where $\textbf{u}$ is the relativistic velocity of the moving bath (more specifically, $\frac{\textbf{v}}{c}$) and $d\textbf{G}$ is the change of momentum due to the exchange of heat. Einstein dubbed this the `translational work'. One can see that when \textbf{u} is 0, the work done reduces to the usual definition. In turn, we generalize the first law from:

\begin{equation}
    dU = dQ - pdV
\end{equation} 

\noindent to:

\begin{equation}
    dU = dQ - pdV + \textbf{u} \cdot d\textbf{G}
\end{equation}

\noindent for a moving system. With this definition of work, and some results from continuum electrodynamics, one can obtain a relationship between the quantities of heat exchanged:\footnote{See Liu (1994, 984 -- 987) for a full derivation.}

\begin{equation}
    \frac{dQ'}{dQ_0} = \frac{1}{\gamma}
\end{equation} 

\noindent and hence, from (11):

\begin{equation}
    T' = \frac{1}{\gamma} T_0
\end{equation}

\noindent where $\gamma$ is the Lorentz factor. We thus arrive at a Lorentz transformation for temperature, according to which a moving system has a lower temperature and appears cooler than a system at rest. This is the \textit{Planck-Einstein formulation} of relativistic thermodynamics.

Secondly, one may, like later Einstein (in private correspondence to von Laue), Ott (1963) and others, doubt the need for translational work. Later Einstein wrote: 

\begin{quote}
    When a heat exchange takes place between a reservoir and a 'machine', both of them are at rest with each other and acceleration-free, it does not require work in this process. This holds independently whether both of them are at rest with respect to the employed coordinate system or in a uniform motion relative to it. (Einstein 1952, quoted in Liu 1994, 199)
\end{quote}

\noindent In the rest frame of the \textit{moving heat bath}, heat exchange is assumed to occur isothermally (as with the usual Carnot cycle) when both the engine and the heat bath are \textit{at rest} with respect to each other. From this perspective, everything should be as they are classically. There should thus be no additional work required other than that resulting from the heat exchange. Put another way, what was thought of as work done to the system in the Planck-Einstein formulation should instead understood as part of heat exchange in the Einstein-Ott-Arzeliés proposal. Without the translational work term in the equation for work, the moving temperature transformation is instead given by:\footnote{See Liu (1992, 197 -- 198) for a detailed derivation.}

\begin{equation}
    T' = {\gamma} T_0
\end{equation}

\noindent and contra (16), the temperature of a moving body appears \textit{hotter}.

I won't pretend to resolve the debate here. However, note that this procedure is not unequivocal on the concept of relativistic temperature: it either appears lower (on the Planck-Einstein formulation) or higher (on the Einstein-Ott formulation) than the rest frame temperature. Importantly, the two proposals reduce to the same classical temperature concept in the rest frame, since the translational work vanishes in this case on both proposals. T$_{classical}$ seems safe, though the fate of its relativistic extension remains undecided.

I end by raising some skepticism about the very idea of a relativistic Carnot cycle, by asking whether there can be a principled answer to whether energy flow is to be understood as `heat' or `work' in such a setting. As Haddad observes, this is problematic due to how energy and momentum are interrelated quantities. A system's energy cannot be uniquely decomposed into heat exchange, internal energy and work: 

\begin{quote}
    In relativistic thermodynamics this decomposition is not covariant since heat exchange is accompanied by momentum flow, and hence, there exist nonunique ways in defining heat and work leading to an ambiguity in the Lorentz transformation of thermal energy and temperature. (Haddad 2017, 39)
\end{quote}

\noindent Since heat flow is accompanied with momentum flow, heat exchange can always be reinterpreted as work done (i.e. as the translational work term).\footnote{See also Dunkel et al (2009, 741).} This raises some initial doubts about the very applicability of thermodynamics beyond the rest frame (i.e. \textbf{CT} in quotidian settings), given the fundamentality of heat and work relations in thermodynamics.

\subsection{The Co-Moving Thermometer: `Moving' Temperature Stays the Same}

The idea that relativistic thermodynamics is essentially `just' quotidian \textbf{CT} is echoed by Landsberg (1970), who builds on the classical concept of a thermometer (and the concept of temperature it establishes) via a \textit{co-moving} thermometer: 

\begin{quote}

One has a box of electronics in both [the relatively moving frame] and [the rest frame] and one arranges, by the operation of buttons and dials to note in [the relatively moving frame] the rest temperature $T_0$ of the system. This makes temperature invariant. (259)

\end{quote}

\noindent The co-moving temperature of any system is stipulated to be its relativistic temperature. But this is \textit{no different than the rest frame temperature of that system}. So the Lorentz transformation according to this procedure is simply

\begin{equation}
    T' = T_0
\end{equation} 

\noindent This proposal can be seen as an extension of T$_{classical}$, in the sense that there is \textit{some} proposed Lorentz transformation. In practice, though, nothing is different from the classical application of a thermometer: we are just measuring the rest frame temperature of the system, as in \textbf{CT}. Landsberg partly justifies this with the claim that ``nobody in his senses will do a thermodynamic calculation in anything but the rest frame of the system". (Landsberg 1970, 260) On this view, \textit{contrary to the relativistic Carnot cycle}, relativistic temperature transforms \textit{as a scalar}, something found in many relativistic thermodynamics textbooks (e.g. Tolman 1934). 

Landsberg (1970, 259) provides an argument for why we couldn't also use this procedure to trivially define alternative Lorentz transformations of other mechanical quantities, e.g. position or time, in terms of their rest frame quantities. He claims that for these mechanical quantities, there \textit{are} measurement discrepancies for the same events in different frames, which needed to be reconciled by Lorentz transformations for consistency. However, for temperature: 

\begin{quote} 
Measurements in a general [reference frame] can be made of mechanical quantities, but in my view not of temperature, [so] our prescription for T' -- namely ``measure $T_0$" -- is quite unsuitable for extension to mechanical quantities. (1970, 259)
\end{quote}

\noindent Prima facie, Landsberg is proposing a novel Lorentz transformation for temperature. However, in my view, this argument amounts to the claim that \textit{there is no relativistic temperature to speak of}; we simply insist on the classical -- rest frame -- temperature concept. His comparison with mechanical quantities makes this clear: the concept of temperature understood via Landsberg's proposal is \textit{not} relativistic the way other quantities are. 

If anything, the preceding discussion suggests that the concept of temperature and its measurement cannot be extended past the rest frame, i.e. into the relativistic domain.\footnote{This is just what physicists do when they consider the temperature of distant astrophysical bodies. They extrapolate and observe other properties of a body -- like luminosity -- associated with its \textit{rest frame} temperature. No consideration of moving temperature is involved.} As Liu (1994, 992) notes: ``The fact seems to be that temperature measurement requires genuine thermal interaction and the state of equilibrium, but when relative macroscopic motion is present, such interaction always disrupts the state of equilibrium and thus renders temperature measurement impossible." Anderson says the same: 

\begin{quote}
    ``Thermodynamic quantities only have meaning in the rest frame of the system being observed. [...] This is not to say that an observer could not infer from measurements on a moving system what its rest temperature is. \textit{The point is that he must interpret these measurements in terms of the rest temperature of the system, since this quantity alone depends on thermodynamic state of the system}." (Anderson 1964, 179 -- 180, emphasis mine) 
\end{quote}

\noindent Repeating Landsberg's words in a different context: that ``nobody in his senses will do a thermodynamic calculation in anything but the rest frame of the system" suggests that the thermodynamic concept of temperature involved here cannot be extended beyond the classical regime. 

\subsection{Relativistic Kinetic Theory: No Fact of the Matter}

An ideal gas can be understood in terms of particles whose velocities are distributed according to the gas's temperature ($\S3.3$). How does \textit{that} notion of temperature extend to relativistic regimes? 

Cubero et al (2007) analyzes the Maxwell-Jüttner distribution, a Maxwell-Boltzmann-type distribution for ideal gases moving at relativistic speeds. They conclude that the temperature should transform as a scalar, i.e. Landsberg's proposal. Interestingly, they explicitly choose a reference frame in which the system is stationary and in equilibrium. But that's just the rest frame of the system! In that case it's unsurprising that there is no transformation required at all for the temperature concept.\footnote{Cubero et al (2007, 3) admits as much when they note that ``Any (relativistic or nonrelativistic) Boltzmann-type equation that gives rise to a universal stationary velocity PDF implicitly assumes the presence of a spatial confinement, thus \textit{singling out a preferred frame of reference}.''}

Elsewhere, Pathria (1966, 794) proposes yet another construction.\footnote{My presentation follows Liu (1994).} They considered a distribution $F$ for an ideal gas in a moving frame with some relativistic velocity $\textbf{u} = \frac{\textbf{v}}{c}$:

\begin{equation}
    F(\textbf{p}) = [e^{(E -\textbf{u} \cdot \textbf{p} - \mu)/kT} + a]^{-1}
\end{equation}

\noindent where \textbf{p} is a molecule's momentum, $E$ its energy, $\mu$ the chemical potential, $k$ the Boltzmann constant, $T$ the system's temperature in that moving frame, and $a$ is 1 or -1 for bosonic and fermionic gases respectively. The distribution then tells us, as with the classical case, how many particles we expect to see with momentum $\textbf{p}$. $F$ is shown to be Lorentz-invariant, and we can compare them as such: 

\begin{equation}
    \frac{E -\textbf{u} \cdot \textbf{p} - \mu}{kT} = \frac{E_0 - \mu_0}{kT_0}
\end{equation}

\noindent With the (known) Lorentz transformations for energy and momentum, we can then show that $T = \frac{1}{\gamma} T_0$, i.e. the \textit{Planck-Einstein} formulation. 

One might think that this suggests some consilience between the kinetic theory and the relativistic Carnot cycle \textit{for} the Planck-Einstein formulation. However, one would be disappointed. Balescu (1968) showed that Parthria's proposed distrbution (20) can be generalized as: 

\begin{equation}
    F^*(\textbf{p}) = [e^{\alpha(\textbf{u})(E -\textbf{u} \cdot \textbf{p} - \frac{\mu}{\beta(\textbf{u)}})/kT} + a]^{-1}
\end{equation}

\noindent with the only constraint that $\alpha(\textbf{0})$ and $\beta(\textbf{0})$ = 1 for arbitrary even functions $\alpha$ and $\beta$. $F^*$ tells us the particle number (or, in quantum mechanical terms, occupation number) associated with some \textbf{p} or $E$ over an interval of time. Balescu shows that any such distribution recovers the usual Maxwell-Boltzmann-type statistics, in the sense that distributions with arbitrary choices of these functions all \textit{agree} on the internal energy and momenta in the rest frame when $\textbf{u} = 0$: T$_{classical}$ is safe from these concerns.

Choosing these functions amounts to choosing some velocity-dependent scaling for temperature via $\alpha$ and chemical potential via $\beta$. Importantly, the question of how temperature scales when moving relativistically is precisely what we want to decide on, yet it is also the quantity rendered arbitrary by this generalization! In particular, Balescu shows that:

\begin{enumerate}
    \item The choice $\alpha$ = 1 amounts to choosing the Planck-Einstein formulation $T = \frac{1}{\gamma} T_0$,
    \item The choice $\alpha$ = $\gamma^2$ amounts to choosing the Einstein-Ott-Arzeliés formulation $T = \gamma T_0$,
    \item The choice $\alpha = \gamma$ amounts to choosing Landsberg's formulation $T = T_0$.
\end{enumerate}

\noindent As Balescu notes: ``Within strict equilibrium thermodynamics, there remains an arbitrariness in comparing the systems of units used by different Lorentz observers in measuring free energy and temperature" and that ``equilibrium statistical mechanics cannot by itself give a unique answer in the present state of development." (1968, 331) Any such choice will be a \textit{postulate}, not something to be assured by the statistical considerations here. In other words, contrary to the classical Maxwell-Boltzmann case, there appears to be no fact of the matter how temperature will behave relativistically, given the underlying particle mechanics. 

Contrary to the classical kinetic theory of heat, which provided a unequivocal conceptual picture (and putative reduction) of T$_{classical}$, there's again no such univocality here.

\subsection{Black-Body Radiation: No Thermality for Moving Black-Bodies}

Finally, when we consider \textit{moving} black-bodies, there is again no clear verdict on the Lorentz transformation for the relativistic temperature. The very concept of a black-body appears to be restricted to the rest frame.

McDonald (2020) provides a simple example of why this is so: consider some observed Planckian (thermal) spectrum of wavelengths from some distant astrophysical object with a peak wavelength $\lambda_{peak}$. We want to ascribe some temperature to that object directly. In our rest frame, using Wien's law (9):

\begin{equation}
    \lambda_{peak} = \frac{b}{T}
\end{equation}

\noindent where $b$ is Wien's displacement constant. Supposing we know the velocity \textbf{v} of the distant astrophysical object, we can compare wavelengths over distances in relativity using the relativistic Doppler effect to find the peak wavelength of the object $\lambda_{peak}'$ at the source:

\begin{equation}
    \lambda_{peak}' = \frac{\lambda_{peak}}{\gamma(1 + \frac{vcos\theta}{c})}
\end{equation}

\noindent where $\theta$ is the angle in the rest frame of the observer between the direction of \textbf{v} and the line of sight between the observer and the object. Given this, we can compare temperatures: 

\begin{equation}
    T' = \frac{\lambda_{peak}}{\lambda_{peak}'}T = \gamma(1 + \frac{vcos\theta}{c})T
\end{equation}
 
\noindent The predicted temperature thus depends on the \textit{direction} of the moving black-body to the inertial observer. 

Landsberg $\&$ Matsas (1996) shows similar results and demonstrates how a relatively moving black-body generally does not have a black-body spectrum from the perspective of an inertial observer. Crucially, they emphasize just how problematic this is for the notion of \textit{black-body radiation} which is defined as \textit{isotropic}: 

\begin{quote}
    [the equation for a moving black-body] cannot be associated with a legitimate thermal bath (which is necessarily isotropic) [...] the temperature concept of a black body is unavoidably associated with the Planckian thermal spectrum, and because a bath which is thermal in an inertial frame $S$ is non-thermal in [a relatively moving] inertial frame $S'$, which moves with some velocity v $\neq$ 0 with respect to S, a universal relativistic temperature transformation [...] cannot exist. (1996, 402--403)
\end{quote}

\noindent In a follow-up article, they further emphasize that ``a moving observer in a heat reservoir can therefore not detect a black-body spectrum, and hence cannot find a parameter which can be identified as \textit{temperature}." (2004, 93) 

The general lesson is simple yet profound. A black-body was defined in the rest frame, i.e. in the non-relativistic setting: we see isotropic radiation with a spectrum, which can be understood to be in equilibrium with other objects and measured as such with thermometers. However, there was no guarantee that a \textit{moving} black-body would still be observed as possessing some \textit{black-body} spectrum with which to ascribe temperature. And it turns out that it generally does \textit{not}. Without this assurance, we cannot reliably use the classical theory of black-body radiation to find a relativistic generalization of temperature.

\section{T$_{classical}$ Falls Apart. What Then?}

Examining four relativistic counterparts to classical procedures thus reveals a \textit{discordant} concept: a moving body may appear to be cooler, or hotter, the same, or may not even appear to be thermal at all. Despite how well these procedures worked classically, they do \textit{not} work together to establish a unequivocal concept of relativistic temperature. Furthermore, \textit{within} each procedure, various conceptual difficulties suggests that the concept of relativistic temperature does not find firm footing \textit{either}. Returning to Einstein's quote, it appears that there is no `natural' way to extend T$_{classical}$.

T$_{classical}$ thus fails to be extended to relativity: well-understood procedures that unequivocally establish its physical meaning in classical settings fail to do so in relativistic settings. These procedures appear to work \textit{just fine} in classical settings, i.e. in the rest frame. However, attempting to extend them to relativistic settings immediately led to conceptual difficulties. This all suggests that the concept of temperature -- and correspondingly, heat -- is inherently a concept restricted to the rest frame. 

More generally, any relativistic extension of \textbf{CT} violates some classical intuitions and will appear `unnatural'. No matter our choice of temperature transformation, something from \textbf{CT} must go. Broadening Balescu's point ($\S4.3$), Landsberg (1970, 263--265) generalizes the thermodynamic relations in terms of arbitrary functions $\theta(\gamma)$ and $f(\gamma)$: 

\begin{equation}
    TdS = \theta dQ
\end{equation}

\begin{equation}
    dQ = fdQ_0
\end{equation}

\noindent where $f$ is the force function:

\begin{equation}
    f = \frac{1}{\gamma} + r(1 - \frac{1}{\gamma^2})
\end{equation}

\noindent where $r = 0$ if we demand the Planck-Einstein translational work, or $r = \gamma$ for the Einstein-Ott view without such work. Again, we only require $\theta(1) = f(1) = 1$ so that in the rest frame everything reduces to \textbf{CT}. Different choices, again, entail different concepts of relativistic temperature, but also other thermodynamic relations (and hence the thermodynamic laws). (See Figure 4.)
\begin{figure}
\begin{center}
    \begin{tabular}{ |p{3.60cm}||p{2cm}|p{2cm}|p{0.70cm}|p{1.75cm}|}
 \hline
 Moving temperature... & $dQ/T dS = \theta $& $dQ/dQ_0 = f$ & $r$ & $T/T_0 = \theta f$ \\
 \hline
 ... is lower & 1 & $1 /\gamma $ & 0 & $1 /\gamma $\\
 ... is higher & 1  & $\gamma$ & $\gamma$ & $\gamma$ \\
 ... is invariant & $\gamma$ & $1 /\gamma $ & 0 & 1\\
 \hline
 \end{tabular}
 
\caption{Figure 4. A list of some choices of $\theta$, $f$, and $r$.}
\end{center}
\end{figure}

Importantly, no choice preserves all intuitions about T$_{classical}$ and \textbf{CT}. Demanding that a lower (higher) moving temperature leads to non-classical behavior. Landsberg (1970, 260 -- 262) considers two thermally interacting bodies $A$ and $B$ moving relatively to one another. $A$ (in its rest frame) sees the other as cooler (warmer) and hence heat flows from (to) $B$. But the same analysis occurs in $B$'s rest frame to opposite effect! So heat flow becomes frame-dependent and indeterminate, contrary to our classical intuitions.

However, demanding temperature-invariance entails that the classical laws of thermodynamics are no longer form-invariant in all inertial frames. Notably, we must revise their form by including some variations of functions $f$, $\gamma$ and $\theta$.\footnote{See Landsberg (1970, 264).} So we preserve some intuitions about heat flow but give up the cherished form of classical thermodynamical laws. Interestingly, it is precisely this classical form that Bekenstein (1973) appealed to, when making the formal analogies between thermodynamics and black holes.

What then? I end with two possible interpretations of my analysis: an \textit{eliminativist} viewpoint, and a \textit{pluralist} viewpoint.\footnote{See Taylor and Vickers (2017) for discussion of this dichotomy.} On the former, one might interpret temperature akin to \textit{simultaneity}: both concepts are well-defined within some rest frame, but there is no absolute fact of the matter as to how they apply \textit{beyond} for relatively moving observers. If one believes that the only physically significant quantities are those which are frame-invariant or co-variant (recall fn. 8), temperature's frame-dependence might lead one to abandon talk of temperature as physically significant, just as we have for simultaneity.\footnote{For discussion of the status of simultaneity, see Janis (2018) and references therein.} 

On the latter, one might \textit{instead} interpret temperature akin to relativistic \textit{rotation}. Analogously, Malament (2000) identifies two equally plausible criteria for defining rotation which agree in classical settings, yet disagree in general-relativistic settings. Importantly, both violate some classical intuitions. Nevertheless: 

\begin{quote}
    There is no suggestion here that [this] poses a deep interpretive problem [...] nor that we have to give up talk about rotation in general relativity. The point is just that [...] we may have to disambiguate different criteria of rotation, and [...] that they all leave our classical intuitions far behind. (2000, 28)
\end{quote}

\noindent Likewise, on this view, we might accept that T$_{classical}$ breaks down, and that (relativistic extensions of) classical procedures fail to unequivocally define a relativistic temperature. However, we need not abandon \textit{temperature} altogether; instead, we need only to work harder to disambiguate and generalize the concept of temperature (and thermodynamical laws).

Depending on interpretation, questions arise. For instance, should formal analogies between black holes and classical thermodynamical laws be taken seriously, if the form of the classical thermodynamical laws doesn't actually survive in relativistic domains? Could typical black holes be treated as `at rest', such that T$_{classical}$ might still apply? Should we, and how should we, generalize T$_{classical}$? I leave these questions to future work.

\section{Conclusion}

The conclusion that classical thermodynamical concepts fall apart in new regimes should not be surprising to philosophers of science. For instance, Callender (2001) cautioned against ``taking thermodynamics too seriously" even in the statistical mechanical regime. He argues that taking the laws and concepts of classical thermodynamics too literally when attempting to formulate a reduction of classical thermodynamics to statistical mechanics leads to error. Furthermore, Callender briefly notes how, similarly, ``perhaps the principal reason for the confusion [in relativizing temperature] is the fact that investigators simply assumed that relativistic counterparts of some laws of thermodynamics would look just like the phenomenological laws -- they took (some) thermodynamics too seriously." (2001, 551) 

Earman (1978, 178) says essentially the same when he diagnoses the problem with relativistic thermodynamics: the pioneers of relativistic thermodynamics acted ``as if thermodynamics were a self- contained subject, existing independently of any statistical mechanical interpretation. Within this setting, many different 'transformation laws' for the thermodynamical quantities are possible." However, the problem is somewhat worse if what I've said in $\S4.3$ is right: even the statistical mechanical determination of a relativistic temperature is up for grabs. 

In any case, I hope to have highlighted how \textit{messy} the situation is in relativistic thermodynamics. Yet, while physicists continue to chime in,\footnote{See McDonald (2020) and references therein.} not much has been said by contemporary philosophers, despite ``how rich a mine this area is for philosophy of science". (Earman 1978, 157) Besides Earman, the only other philosopher to have discussed this topic in detail appears to be his student, Liu. (1992 and 1994) Through this paper, I hope to have at least re-ignited some interest in this topic.

\vspace{1.2cm}



\defbibheading{myheading}[\footnotesize{REFERENCES}]{%
\centering \normalfont\textsc{{#1}}%
}

\printbibliography[heading=myheading]

\nocite{*}

\end{document}